\documentstyle[prl,aps,epsfig,multicol]{revtex}

\renewcommand{\narrowtext}{\begin{multicols}{2} \global\columnwidth20.5pc}
\renewcommand{\widetext}{\end{multicols} \global\columnwidth42.5pc}
\newcommand{\Lrule}{\vspace*{-0.2in}\noindent\vrule width3.5in height.2pt
  depth.2pt \vrule depth0em height1em}
\newcommand{\Rrule}
{\vspace*{-0.05in}
\hfill\vrule depth1em height0pt \vrule width3.5in height.2pt depth.2pt\vspace*{-1pc}}
\begin{document}

\title{Many-body spin Berry phases emerging from the $\pi$-flux state: 
antiferromagnetic/valence-bond-solid competition}

\author{ 
Akihiro Tanaka and Xiao Hu
}
\address{Computational Materials Science Center, National Institute for Materials Science, 
Sengen 1-2-1 Tsukuba 305-0047, Japan}

\date{}

\maketitle

\begin{abstract}  
We uncover new topology-related features of the $\pi$-flux saddle-point solution of the 
$D$=2+1 Heisenberg antiferromagnet.   
We note that symmetries of the spinons  
sustain a built-in competition between antiferromagnetic (AF) and valence-bond-solid (VBS) orders, 
the two tendencies central to recent developments on quantum criticality.   
An effective theory     
containing an analogue  
of the %well-known $D=1+1$ 
Wess-Zumino-Witten term is derived,  
which generates quantum phases related to  
AF monopoles with VBS cores, and reproduces Haldane's hedgehog 
Berry phases. 
The theory readily generalizes to $\pi$-flux states for all $D$.

\end{abstract}

%\pacs{PACS number: 13.20.He, 11.30.Er, 12.15Hh, 11.80Et}

%\maketitle

\begin{multicols}{2}[]

\noindent

Our understanding on quantum critical points\cite{Senthil}/phases\cite{Hermele} in $D$=2+1 antiferromagets, and the issue of deconfinement therein 
have recently undergone a rapid sequence of 
developments. Competition between  
antiferromagnetic (AF) and valence-bond-solid (VBS)-like fluctuations 
constitute the basic premises for much of these activities. 
%Building on earlier works of 
%Haldane\cite{Haldane} and Read and Sachdev\cite{Read-Sachdev}, 
These theories have brought into wide recognition the relevance of monopole defects of 
the AF order-parameter and in particular      
the nontrivial Berry phase factors\cite{Haldane,Read-Sachdev} associated with such objects. 
Here, with these new perspectives, 
we revisit the Berry phase effect\cite{Marston} in states emerging from 
the $\pi$-flux saddle point solution 
of the Heisenberg antiferromagnet\cite{Affleck Marston}, 
a popular point of departure for studying undoped and lightly doped cuprate 
Mott insulators. We find that their topological properties are 
rather rich. 
Among our findings are (1) a {\it chiral} symmetry of the 
$\pi$-flux Dirac fermion 
relating the AF and VBS orders, which lead us to a natural framework 
for studying their mutual competition, (2) a low energy effective theory 
with a novel many-spin Berry phase term for which the contributions from a 
${\it composite}$ defect (see below) reproduce the monopole 
Berry phases, 
(3) a natural extension of such framework to arbitrary dimensions with 
possible relevance to higher dimensional spin liquids. 

It is worth digressing on the second point before 
proceeding to the more technical aspects. 
An important feature of monopole excitations is the energy cost due to the 
rapid modulation of the AF order near the singular core. 
Meanwhile, in the discussions which follow,  
the system takes advantage of the inherent AF-VBS competition and saves energy 
by escaping into a local VBS state at the defect cores. 
Such physics share in spirit with work by  
Levin and Senthil\cite{Levin Senthil}, who study AF-VBS competition starting from the 
VBS side. In that work, the four-state clock ordering of the VBS state   
is disordered through the introduction of $Z_4$ vortices.  
Close inspection of the lattice model shows that these defects have an AF 
core, as opposed to conventional vortices with featureless singular cores; 
hence their condensation leads to the N{\'e}el state. Likewise, it is natural to expect a 
VBS core to be present in a hedghog-like configuration of the AF order parameter, the condensation 
of which would give way to a VBS state.  
Indeed we will see that incorporation of this feature is essential 
in recovering Haldane's Berry phase starting from the $\pi$-flux state.   

\noindent{\it Continuum fermion model}
The $\pi$-flux hopping Hamiltonian on a two dimensional square lattice is 
${\cal H}_{\pi}=\sum_{i,\mu, \sigma}t c_{i\sigma}^{\dagger}{\bf T}_{\mu}c_{i\sigma}$, where 
${\bf T}_{\mu}$ with $\mu$=x, y generates translation by one site. 
The $\pi$-flux condition imposes the anticommutation relation $\{{\bf T}_{x}, {\bf T}_{y}\}=0$, which immediately leads to the spinon's 
dispersion $E({\mbox {\boldmath $k$}})=\pm t{\sqrt{\cos^2 k_x +\cos^2 k_y}}$ with Dirac nodes at 
${\mbox {\boldmath $k$}}=(\frac{\pi}{2}, \pm\frac{\pi}{2})$. 
It is convenient to group together the four cites sharing 
a unit plaquette (Fig.1) into components (with spin indices) 
of a Dirac spinor, 
${\Psi}=^{t}(\psi_{1\sigma}, \psi_{2\sigma}, \psi_{3\sigma}, \psi_{4\sigma})$
\cite{Ludwig}.

%\vspace*{-5mm}
%%\vspace*{15mm}
%\begin{figure}[h]
%\epsfxsize 7.8cm
%%\epsfxsize 3.0cm
%%\centerline {\epsfbox{unit_cell1.eps}}
%\centerline {\epsfbox{cell2.eps}}
%\caption{lattice used to derive continuum Dirac theory}
%\end{figure}

\vspace*{15mm}
\begin{figure}
\vspace{3cm}
\includegraphics{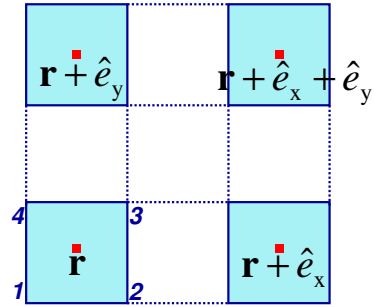}
\caption{Lattice used to derive continuum Dirac theory}
\end{figure}

To fix the representation of the Dirac gamma matrices,  
we account for the  
$\pi$-flux condition by  
assigning negative hopping integrals $-t$ to 
links residing on every other horizontal rows; all other links have positive hopping integrals, $+t$. 
Linearizing around the nodes, we arrive at the Dirac action (hereafter we employ Euclidean spacetime conventions) 
${\cal L}=i{\bar \Psi}{\not\! \partial}\Psi$, where the slash indicates the contraction with the gamma matrices  
$\gamma_{0}=\sigma_{0}\otimes\sigma_{z}, \gamma_{1}
=-\sigma_{0}\otimes\sigma_{x}, 
\gamma_{2}=\sigma_{y}\otimes\sigma_{y}$. Here the first (second) matrix 
within a direct product determines the block structure (the matrix elements within the cells). 

Hermele {\it et al} have recently shown 
that the algebraic spin liquid described by the $\pi$-flux Hamiltonian is stable 
against monopoles, at least for large-$N$\cite{Hermele}. 
We now move away from this critical phase by supplementing the 
theory with mass terms so that the system can acquire AF order.   
Indeed, earlier works on the $\pi$-flux state show\cite{Hsu} that a substantial improvement on the 
variational energy is acheived by adding on a spin density wave term 
${\cal H}_{\rm SDW}=\sum_{i}M (-1)^{i_{x}+i_{y}}c_{i\sigma}^{\dagger}\sigma_{z}c_{i\sigma}$, 
which appears to be in accord with angular resolved 
photoemission experiments on cuprate Mott insulators\cite{Ronning}. 
Being interested in extracting the dependence of the effective action 
and possible Berry phase terms on the 
N{\'e}el unit director 
${\mbox {\boldmath $n$}}$, we make in ${\cal H}_{\rm SDW}$ the generalization 
$\sigma_{z}\rightarrow{\mbox{\boldmath $n$}}\cdot{\mbox{\boldmath $\sigma$}}$, 
which is physically a potential energy term imposing the generation of AF spin moment with the 
space-time-dependent orientation ${\mbox {\boldmath$n$}}(\tau, x, y)$. In the continuum limit 
this becomes the mass term  
${\cal L}_{\rm AF}=im_{\rm AF}{\bar \Psi}({\mbox {\boldmath$n$}}\cdot{\mbox{\boldmath $\sigma$}}){\Psi}$.    
Next, we recall\cite{Ludwig,QED_3} that 
in contrast to usual irreducible representations for 
Dirac fermions in $D$=2+1 where 
the notion of chirality is absent (no ^^ ^^ $\gamma_{5}$"s), 
we have $two$ generators of chiral transformations at our disposal, 
$\gamma_{3}=\sigma_{z}\otimes\sigma_{y}$ and $\gamma_{5}=\sigma_{x}\otimes\sigma_{y}$ which anticommute 
mutually as well as with the space-time components $\gamma_{0}, \gamma_{1}$ and $\gamma_{2}$. 
One reads off from the explicit matrix elements that the effects of 
chiral mass terms 
proportional to ${\bar \Psi}\gamma_{3}\Psi$ and 
${\bar \Psi}\gamma_{5}\Psi$ each amount to breaking lattice translational symmetry by 
introducing bond alternations  $t\rightarrow t+(-1)^{i_{\mu}}\delta t$ in the horizontal ($\mu$=x) 
and vertical ($\mu$=y) directions. A crucial observation here is 
that the SDW and the 
two VBS ordering potentials 
$i{\bar \Psi}Q\Psi$, ${\bar \Psi}\gamma_{3}\Psi$ 
and ${\bar \Psi}\gamma_5 \Psi$ 
($Q\equiv{\mbox{\boldmath$n$}}\cdot{\mbox{\boldmath$\sigma$}}$) 
all belong to 
the family of chirally rotated mass terms 
${\cal L}_{\rm chiral}\propto i{\bar \Psi}Q e^{i(\alpha \gamma_{3}+\beta\gamma_{5})\otimes Q}\Psi$ 
($\alpha, \beta\in {\bf R}$), i.e. they 
transform into one another by suitable chiral transformations. 
Finally we mention that in the 4D irreducible representation we 
have another possible mass term \cite{QED_3} 
${\bar \Psi}\gamma_{3}\gamma_{5}\Psi$  
which has relevance to chiral spin liquids (see footnote \cite{note2}). 
We do not retain this P- and T-violating term. 

\noindent
{\it Bosonization} 
%Summarizing the situation so far, (1)
%the Clifford algebra for the Dirac fermion 
%arising from the $\pi$-flux state is spanned by a total of five gamma matrices,% two of which are related to 
%VBS ordering in the $x-$ and $y$-directions,  
%and (2) the horizontal and vertical VBS ordering potentials and the AF SDW pote%ntial rotate into one 
%another by suitable chiral transformations. 
The preceeding implies that the spontaneous breaking of the 
chiral symmetry present in the $\pi$-flux state can lead to 
AF or VBS orders, depending on the chiral angles $\alpha$ and $\beta$. 
This motivates us to study the AF-VBS competition in terms 
of the following theory with a generalized mass term (equivalent to ${\cal L}_{\rm chiral}$),
\begin{equation} 
{\cal L}_{\rm F}^{2+1}=i{\bar \Psi}\left[{\not\! \partial}+mV_{(2+1)}\right]\Psi,         
\end{equation}
where 
%\begin{equation}
$V_{(2+1)}\equiv v^1 \sigma_x +v^2 \sigma_y +v^3 \sigma_z +i\gamma_3 v^4 +
i\gamma_5 v^5$, 
%\label{2+1D V} 
%\end{equation}
and ${\mbox {\boldmath $v$}}_{(2+1)}\equiv(v^{1},...,v^{5})$ is a five component unit vector. 
The first three components 
comprise a vector ${\mbox {\boldmath $v$}}_{\rm AF}\equiv(v^{1}, v^{2}, v^{3})$ which is parallel to 
${\mbox {\boldmath $n$}}$ and in competition with a VBS-like order parameter $(v^{4}, v^{5})$. We now show that 
this theory can be ^^ ^^ bosonized" in terms of ${\mbox {\boldmath$v$}}_{(2+1)}$, to yield an effective action 
which contains a new Berry phase term.  
Central to this feat is the following relation\cite{Abanov,note1} 
satisfied by the Dirac operator 
${\cal D}[{\mbox {\boldmath $v$}}_{(2+1)}]
=i{\not\!\partial}
+imV_{(2+1)}$ and its hermitian conjugate,  
%$D[{\mbox {\boldmath $v$}}_{(2+1)}]^{\dagger}=i{\not\!\partial}
%-im(v_1 \sigma_1 +v_2 \sigma_2 +v_3 \sigma_3 -i\gamma_3 v_4 -
%i\gamma_5 v_5)$ satisfies 
\begin{equation}
{\cal D}^{\dagger}{\cal D}
%&=&-\partial^2+m^2-m{\not\!\partial}\cdot V_{(2+1)} 
%+m\gamma_{\mu}V_{(2+1)}\partial_{\mu}\nonumber \\
= -\partial^2 +m^2 -m{\not\!\partial}V_{(2+1)},
\label{DdaggerD}   
\end{equation}
which enables one to rewrite the  
variation of the fermionic determinant ${\cal S}_{\rm eff}=-{\rm ln}{\rm det}
{\cal D}[{\mbox{\boldmath $v$}_{(2+1)}}]$ 
into a form suitable for generating a derivative expansion: 
\begin{equation} 
\delta S_{\rm eff}
%-trD^{-1}\delta D \nonumber \\
%&=& -tr(D^{\dagger}D)^{-1}D^{\dagger}\delta D \nonumber \\  
= -tr\left[
\frac{1}{-\partial^2 +m^2 -m{\not\!\partial}V_{(2+1)}}
{\cal D}^{\dagger}\delta {\cal D}
\right].
\label{expansion}
\end{equation} 
It is easy to see that a nonlinear sigma (NL$\sigma$) model 
${\cal S}_{{\rm NL}\sigma}=\frac{1}{2g}\int d^3 {\rm x} 
(\partial_{\mu}{\mbox{\boldmath $v$}}_{(2+1)})^2 $ arises, with $g$ a nonuniversal coupling constant. Less trivial is  
an imaginary contribution to eq.(\ref{expansion}), 
$\delta{\cal S}_{\rm BP}^{2+1}$, which we pick up at  
third order in powers of ${\not\!\partial}V_{(2+1)}$. 
As is usual with Wess-Zumino type terms, 
one recovers the action 
${\cal S}_{\rm BP}^{2+1}$ from its variation $\delta{\cal S}_{\rm BP}^{2+1}$ 
with the aid of an auxiliary variable $t\in[0,1]$ which smoothly sweeps  
%With the introduction of $t$, ${\mbox {\boldmath $v$}}_{(2+1)}({\rm x})$ is embedded  
the extended function ${\mbox {\boldmath $v$}}_{(2+1)}(t, x_{\mu})$ 
between its two asymptotics, a     
fixed value at $t=0$, say 
%${\mbox {\boldmath $v$}}_{(2+1)}(t=0, x_{\mu})\equiv(0,0,0,0,1)$, 
(0,0,0,0,1), 
and the physical value at $t=1$, 
%${\mbox {\boldmath $v$}}_{2+1}(t=1, x_{\mu})
%\equiv{\mbox {\boldmath $v$}}_{(2+1)}(x_{\mu})$. 
${\mbox {\boldmath $v$}}_{(2+1)}(x_{\mu})$. 
%In terms of the components of this new vectorial function, the imaginary term can now be expressed as
The result is 
\begin{equation}
{\cal S}_{\rm BP}^{2+1}
\mbox {\hspace*{-1mm}}
= \mbox {\hspace*{-1mm}}
\frac{-2\pi i \epsilon_{abcde}}
{Area(S^4)}\mbox {\hspace*{-1mm}}
\int_{0}^{1} \mbox {\hspace*{-2mm}}dt
\int {\mbox {\hspace*{-2mm}}}d^3{\rm x}
v^{a}\partial_{t}v^{b}\partial_{\tau}v^{c}\partial_{x}v^{d}\partial_{y}v^{e}, 
\label{2+1D BP term}
\end{equation} 
where $Area(S^{4})=\frac{2\pi^{\frac{5}{2}}}{\Gamma(\frac{5}{2})}=\frac{8}{3}\pi^2$. 
Topologically, this is -$i2\pi$ times the  
winding number which counts the number of times 
the compactified ^^ ^^ space-time" $\{(t, x_{\mu})\}$ isomorphic to $S^3\times S^1 
\sim S^4$   
wraps around the target space (also $S^4$) for 
${\mbox{\boldmath $v$}}_{(2+1)}(t, x_{\mu})$. 
It is important not to confuse this action with Hopf or Chern-Simons terms, which have been  
studied extensively in the context of chiral spin systems\cite{Fradkin book}  
and are strongly tied to the dimensionality $D$=2+1.  
Indeed, as we now show, the foregoing readily generalizes to theories of AF-VBS competition 
in arbitrary space-time dimensions $D$=$d$+1, 
where for each $d$ we find topological terms that are generalized versions of ${\cal S}_{\rm BP}^{2+1}$. 
We will return to the physical contents for the specific case of $D$=2+1 later.  

\noindent{\it AF-VBS competition for general D} 
Our detour starts by mentioning a generic property of 
Clifford algebras\cite{Lie algebra dictionary} which lies behind this generalization. 
%
%\noindent
{\sl Property: Let $n$ be the number of matrices spanning the algebra 
$\{\gamma_{i}, \gamma_{j}\}=2\delta_{ij} (i, j=1,...,n)$. Representations for this algebra 
can be realized by a set of $2^{p}\times 2^p$ $\gamma_{i}$-matrices where either $n=2p$ or $n=2p+1$.} 
%
%\noindent
To see why this goes hand in hand with the construction of a fermionic theory describing 
AF-VBS competition let us consider a $\pi$-flux state on a $d$-dimensional 
hypercubic lattice. 
The latter, in analogy with the $D$=2+1 case, is defined by the anticommutivity among the  
generators of translation $\{ {\bf T}_{l}, {\bf T}_{m}\}=0$ for $l\neq m (l, m=1,...,d)$. 
This gives rise to Dirac nodes within the Brillouin zone.(For d=1, where there  
are no notions of flux-lines which pierce 
plaquettes, it suffices to simply start with a free tight-binding model 
which gives rise to massless Dirac fermions. 
The arguments below applies for this case as well.)
In going to the continuum language, Dirac spinors are 
constructed by dividing all lattice sites into cells consisting of $2^d$ sites. 
The Dirac matrices $\gamma_{\mu}$ are therefore $2^d \times 2^d$ matrices. 
Meanwhile, what we wish to construct is a fermionic theory of the form 
${\cal L}_{\rm F}^{d+1}=i{\bar \Psi}[{\not\! \partial}+V_{(d+1)}]\Psi$, 
where notations are obvious extensions from those used in the $D$=2+1 case. 
The number of Dirac matrices required for this purpose is    
$2d+1$, there being in addition to the $d$+1 space-time components $\gamma_{0},...,\gamma_{d}$, 
a total of $d$ chiral matrices (^^ ^^ $\gamma_5$"s),  
each standing for the directions available for dimerization. We see that this fits in nicely with the 
mathematical property forementioned 
when we put $p=d$. (It is also straightword to work out an explicit derivation of ${\cal L}_{\rm F}^{d+1}$ 
starting from the lattice theory.)     

The Dirac operator $D[{\mbox{\boldmath $v$}}_{(d+1)}]$ obeys eqs.(\ref{DdaggerD}) and (\ref{expansion}), in 
which the replacement ${\mbox{\boldmath $v$}}_{(2+1)}\rightarrow{\mbox{\boldmath $v$}}_{(d+1)}$ is to be made. 
Carrying out the derivative expansion as before we obtain the low energy effective theory which 
is an $O(3+d)$ NL$\sigma$ model suppplemented with the 
topological term 
%%%%%%%%%%%%%%%%%%%%%%%%%%%%%%%%%%%%%%%%%%%%%%%%%%%%%%%%%%%%%%%%%%%%%%%%%%%
\widetext
\Lrule
\begin{equation}
{\cal S}_{\rm BP}^{d+1} =
\frac{-2\pi i}
{Area(S^{d+2})}
\int_{0}^{1} dt
\int d^{d+1}{\rm x}\epsilon_{\alpha_{1}\cdot\cdot\cdot\alpha_{d+3}}
v^{\alpha_{1}}\partial_{t}v^{\alpha_{2}}\partial_{\tau}v^{\alpha_{3}}\partial_{x_{1}}v^{\alpha_{4}}
\cdot\cdot\cdot\partial_{x_{d}}v^{\alpha_{d+3}}.
\end{equation} 
\Rrule
\narrowtext\noindent
%%%%%%%%%%%%%%%%%%%%%%%%%%%%%%%%%%%%%%%%%%%%%%%%%%%%%%%%%%%%%%%%%%%%%%%%%%%%
%\begin{eqnarray}
%{\cal S}_{\rm BP}^{d+1} &=&
%\frac{-2\pi i}
%{Area(S^{d+2})}
%\int_{0}^{1} dt
%\int d^{d+1}{\rm x}\epsilon^{\alpha_{1}\cdot\cdot\cdot\alpha_{d+3}}
%\nonumber \\ 
%&&v^{\alpha_{1}}\partial_{t}v^{\alpha_{2}}
%\partial_{\tau}v^{\alpha_{3}}\partial_{x_{1}}v^{\alpha_{4}}
%\cdot\cdot\cdot\partial_{x_{d}}v^{\alpha_{d+3}}.
%\end{eqnarray} 
%
For $D$=1+1 ($d$=1), the isotropic $O$(4) theory 
%which derives from 
with 
the partition function 
$Z[{\mbox{\boldmath$v$}}_{(1+1)}]=\int D{\mbox{\boldmath$v$}}_{(1+1)} e^{-({\cal S}_{{\rm NL}\sigma}+{\cal S}_{\rm BP}^{1+1})}$ 
has been identified (e.g. \cite{Abanov,Tanaka}) with the conformally 
invariant $SU(2)_{1}$ Wess-Zumino-Witten (WZW) model, 
the fixed point theory for the $S$=1/2 Heisenberg antiferromagnet\cite{Affleck Haldane}.   
It is instructive to analyze the effect of introducing different types of 
anisotropy between the AF and dimer sectors in this model\cite{Tanaka}, 
breaking down the symmetry to 
$O(3)\times Z_{2}$ or lower (the appearance of $Z_2$ is a remnant of the 
underlying lattice).  
First, the effective theory for the AF limit reduces at the semiclassical level 
to the $O$(3) NL$\sigma$ model at 
topological angle $\theta=\pi$, as originally 
proposed by Haldane\cite{Haldane}. 
Taking the opposite limit with complete dimerization   
makes the kinetic and Berry phase terms of the $O$(3) AF sector vanish, 
reflecting the quenching of the spin moment. 
An intermediate situation arises when the anisotropy modulates in space, 
physically corresponding to a distribution of 
nonmagnetic impurities. This induces 
$S$=1/2 moments in the background of a singlet state, 
whose spin Berry phases 
are responsible for novel power-law correlations. 
The isotropic (WZW) point may be 
viewed in light of this picture as the case where 
the anisotropy acquires a temporal dependence as well. 
The main insight gained from these examples  
is how   
the interplay between AF (spin moment generating) and dimer (spin moment quenching) 
ordering tendencies determines the Berry phase, which in turn acts back on the 
ordering of the system. Turning to higher dimensions,  
three dimensional spin liquid systems which may be realized in frustrated 
magnets have lately received considerable interest, where again subtle Berry phase effects due to monopole configurations 
can be present\cite{d=3 spin liquids}.  
We believe our approach as applied to the $D$=3+1 case 
with the novel Berry phase term ${\cal S}_{\rm BP}^{3+1}$ 
provides a new route to capture the topological properties of such systems.\\
\noindent{\it Monopoles and Berry phases}
Returning now to the $D$=2+1 case, the isotropic $O$(5) theory has the interesting feauture of being 
the higher dimensional analogue of the WZW model in the sense detailed above. This 
parallelism has been explored 
to some extent for related models within the context of $D$=2+1 quantum chromodynamics\cite{QCD_3}, where  
current algebras similar to those which determine 
the conformal field theory contents 
of the WZW model were derived. 
We need be aware though, of 
the privellaged role of dimensionality $D$=1+1 for which
the coupling constant becomes dimensionless. Hence the intriguing possibility 
of a nontrivial infrared fixed point is unresolved, and warrants 
further study from the viewpoint of exotic quantum spin systems. \\
We now let an anisotropic term favoring the AF sector, e.g. of the form 
$-\alpha{{\mbox {\boldmath$v$}}_{\rm AF}}^2$ with $\alpha>0$, 
take us away from the isotropic regime. In contrast to  
similar models in the {\it irreducible} representation\cite{Jaroszewicz}, here 
there are no topologically conserved fermionic currents which forbids changes in the Skyrmion number 
${\cal Q}_{xy}=\frac{1}{4\pi}\int dxdy {\mbox{\boldmath $n$}}\cdot
\partial_{x}{\mbox{\boldmath$n$}}\times\partial_{y}{\mbox{\boldmath$n$}}$\cite{note2}.   
Finding Berry phases accompanying such processes 
requires us to extract the dependence of eq.(\ref{2+1D BP term}) on ${\mbox{\boldmath$n$}}$, which 
proceeds in two steps.     
We first integrate over the auxiliary variable $t$. We use without loss of generality 
the parametrization 
%(which satisfy the constraints imposed at $t$=0 and $t$=1) 
$v^{1}=\sin(t\varphi){\pi}^{1}$,  
$v^{2}=\sin(t\varphi){\pi}^{2}$, 
$v^{3}=\sin(t\varphi){\pi}^{3}$, 
$v^{4}=\sin(t\varphi){\pi}^{4}$, and 
$v^{5}=-\cos(t\varphi)$, where ${\mbox{\boldmath$\pi$}}=(\pi^{1},..,\pi^{4})$ is a four component 
unit vector. The resulting Lagrangian density is
%\begin{eqnarray}
%{\cal L}&=& \frac{1}{2g}\left[\sin^2 \varphi(\partial_{\mu}{\mbox{\boldmath$\pi%$}})^2 
% + (\partial_{\mu}\varphi)^2 \right]\nonumber \\
% &+& i\theta q_{\tau xy} +{\cal L}_{\rm anis},  
%\end{eqnarray}
\begin{equation}
{\cal L}=\frac{1}{2g}
\left[\sin^2 \varphi(\partial_{\mu}{\mbox{\boldmath$\pi$}})^2 
 + (\partial_{\mu}\varphi)^2 \right]
 + i\theta q_{\tau xy} +{\cal L}_{\rm anis}, 
\end{equation}
where $\theta=\pi(1-\frac{9}{8}\cos\varphi+\frac{1}{8}\cos 3\varphi)$,  
$q_{\tau xy}\equiv \frac{1}{2\pi^2}\epsilon_{abcd}\pi^a \partial_{\tau}\pi^b 
\partial_{x}\pi^c 
\partial_{y}\pi^d$, and ${\cal L}_{\rm anis}$ is the anisotropy term. 
Notice that the first and third terms vanish, as they must, under the 
spin-moment quenching condition $\varphi=0$. 
When the phase field $\varphi$ is locked at a constant value 
-corresponding to fixing 
the bond alternation strength in the vertical direction 
relative to the magnitude of the remaining four components 
of ${\mbox{\boldmath$v$}}$, 
this reduces to the $D$=2+1   
$O(4)$ nonlinear sigma model with a $\theta$-term\cite{Abanov}. 
(This intermediate model should thus be relevant to 
spin systems with anisotropic bond alternation\cite{Dombre,Katoh}.) 
%
%\begin{figure}[h]
%\epsfxsize 8.0cm
%\centerline {\epsfbox{z4.eps}}
%\caption{A sequence of $\frac{\pi}{2}$ rotations around a dual site 
%which simulateously rotates the direct sites and the 
%VBS order parameter.}
%\end{figure}
%
Going on to the second step, we now parametrize the components of 
${\mbox{\boldmath $\pi$}}$ as 
$\pi^{1}=\sin\phi n^1$, 
$\pi^{2}=\sin\phi n^2$,
$\pi^{3}=\sin\phi n^3$,
$\pi^{4}=-\cos\phi $. The Berry phase term can now be recasted in a way 
which explicitly depends on monopole-like configurations
\cite{Sachdev Jalabert}. The result, obtained 
by integrating by parts, is
\begin{equation}
{\cal L}_{\rm BP}=-\frac{i}{2}(2\phi-\sin2\phi)
(1-\frac{9}{8}\cos\varphi +\frac{1}{8}\cos 3\varphi)\rho_{m},
\label{monopole BP}
\end{equation} 
where $\rho_{m}=\partial_{\tau}(\frac{\epsilon_{abc}}{4\pi}
n^a \partial_{x}n^b \partial_{c}n^c )$+$c.p.$ 
is the monopole charge density. 
Integrating $\rho_m$ over a space-time region surrounding the 
center of the monopole event gives the 
change in the Skyrmion number 
between the two time-slices 
before and after occurence of the event, i.e. 
$\int d\tau dxdy \rho_{m}=\Delta {\cal Q}_{xy}$. 
%\begin{figure}[h]
%\vspace*{-5mm}
%\epsfxsize 9cm
%\centerline {\epsfbox{z4.eps}}
%\caption{A sequence of $\frac{\pi}{2}$ rotations around a dual site 
%which simulateously rotates the direct sites and the 
%VBS order parameter.}
%\end{figure}

%\vspace*{35mm}
\begin{figure}
\vspace{3cm}
\includegraphics{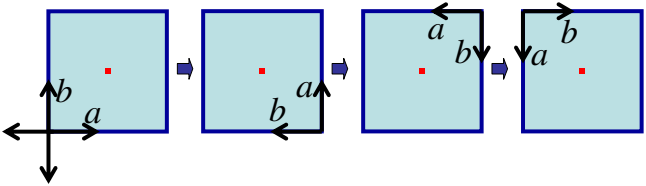}
\caption{Sequence of $\frac{\pi}{2}$ rotations around a dual site 
which simulateously rotates the direct sites and the 
VBS order parameter.}
\end{figure}

A spatially modulated pattern in the 
monopole Berry phase \cite{Haldane,Read-Sachdev}
arises from this term in the following way. At 
each lattice site there is a competition between spin momentum generation and 
a local $Z_4$-valued VBS order. While the bulk favors the former due to 
the presence of ${\cal L}_{\rm anis}$, the latter emerges locally when a 
monopole happens to be centered at that particular site. 
We may choose this VBS core to be represented e.g. at sublattice 1 in Fig. 1 
by the combination 
$\varphi=\frac{\pi}{2}$ and $\phi=0$, which implies, according to 
eq.(\ref{monopole BP}) that ${\cal S}_{\rm BP}^{1}=1$, with the superscript 
standing for the sublattice index. We then go around the plaquette 
counterclockwise as  depicted in Fig. 2, which simultaneously rotates the 
orientation of the VBS order parameter by 90 degrees increment. Noting 
that the orientation of the ^^ ^^ VBS clock" 
is specified by the angle $\phi-\varphi$, we must 
correct for this by also incrementing $\phi$ by $-\frac{\pi}{2}$ 
(while keeping $\varphi$ fixed)or $\phi$ by $\frac{\pi}{2}$ (keeping 
$\varphi$ fixed). Either way the Berry phase shifts by 
$\frac{\pi}{2}\Delta {\cal Q}_{xy}$. In this way, we find that 
in order to have AF monopoles with VBS cores having the 
same orientation for all four sites sharing the plaquette, 
we must have  
${\cal S}_{\rm BP}^2 =e^{i\frac{\pi}{2}\Delta {\cal Q}_{xy}}$, 
${\cal S}_{\rm BP}^3 =e^{i\pi\Delta {\cal Q}_{xy}}$, 
${\cal S}_{\rm BP}^4 =e^{i\frac{3\pi}{2}\Delta {\cal Q}_{xy}}$. 

This  
methods also applies to the case  
where the VBS state is favored in the bulk, where one recovers the  
Berry phase for the AF core in the VBS vortex
\cite{Levin Senthil,Tai Kai}, 
$\frac{1}{2}(-1)^{i_{x}+i_{y}}\omega$, where $\omega$ is the solid angle 
subtended by the spin. In addition, the framework can be extended to make it applicable 
for studying {\it staggered} flux states. 
%We finally mention that with some sophistication, similar 
%analysis can also be made for the AF on a $D$=2+1 honeycomb lattice
%\cite{Read-Sachdev}. 
In this sense, our method is capable of 'generating' 
a rich variety of spin Berry phase effects. 

In summary we have shown that the $\pi$-flux state perturbed by 
competing AF and VBS orders provides a natural framework for studying 
effective theories which incorporates defects and their Berry phases 
in $D$=2+1 antiferromagnets. We have also demonstrated that the same 
methods also offers a framework to explore     
exotic Berry phases in $D$=3+1 spin systems as well. 

We acknowledge helpful discussions and/or correspondences with Profs. 
A. Abanov, L. Balents, G. Baskaran, I. Herbut, C. Mudry, S. Murakami 
and L. Yu.

\end{multicols}

\begin{thebibliography}{References:}
\bibitem{note1}It is interesting to note 
the difference of our $V$ with that given in  
ref\cite{Abanov} (see eq.(18) of this paper), 
which stems from our use of the 4$D$ representation while 
the 2$D$ representation is employed in ref\cite{Abanov}. 
Indeed our fermionic action can be recasted into a model in the 
latter representation with an internal $SO$(5) symmetry. 
We are indebted to A. Abanov for kindly clarifying this point.
\bibitem{note2} Instead, one is lead to an interesting relation  
$<{\bar \Psi}\gamma_{3}\gamma_{5}\gamma_{0}\Psi>\propto q_{xy}
$ where $q_{xy}$ is the Skyrmion number density 
(i.e. the local spin chirality).
\end{thebibliography}
\end{document}